\documentclass[11pt]{article}

\usepackage{subcaption}
\usepackage{rotating}
\usepackage{pgfplots}
\usepackage{tikz}
\usetikzlibrary{arrows.meta}
\usepackage{enumerate}
\usepackage[utf8]{inputenc}
\usepackage[T1]{fontenc}
\usepackage{lmodern}
\usepackage[english]{babel}
\usepackage{amsmath, amssymb, amsthm, mathtools}
\usepackage{algorithm}
\usepackage{algpseudocode}
\DeclareMathOperator*{\argmin}{arg\,min}
\DeclareMathOperator*{\argmax}{arg\,max}

\DeclareMathOperator{\dist}{d}
\usepackage{graphicx}
\usepackage{geometry}
\usepackage{dcolumn}
\usepackage[hidelinks]{hyperref}
\usepackage{cite}
\usepackage{authblk}
\usepackage{booktabs}
\usepackage{xcolor}
\newtheorem{definition}{Definition}

\usepackage{soul}
\newcommand{\midi}{m}

\date{}

\title{Tonnetz-Driven Graph Wedgelet for Harmonic Complexity Reduction in Music Scores}

\author[1]{Emmanuel Caronna}
\author[1]{Elisa Francomano\thanks{Corresponding author: \texttt{elisa.francomano@unipa.it}}}
\author[1]{Silvia Licciardi}

\affil[1]{Department of Engineering, University of Palermo,
Viale delle Scienze, 90128 Palermo, Italy}

\begin{document}
\maketitle

\begin{abstract}
\noindent
Heterogeneous graph built on notes, lyric syllables, and accompaniment events is a natural representation of symbolic music score, providing
a substrate for both philological analysis and computational tasks. Music features are therefore well-captured by graph geometry and its properties. This representation has proved effective for analytical tasks as cadence detection, voice separation, and stylistic classification. In the present work, the reduction of harmonic complexity of a music score on graph, by preserving task-relevant information, relation between notes, and graph structure is investigated.
A compression scheme for the
piano subgraph of vocal-pianistic scores, built on binary wedge partitioning trees, is proposed. The wedges are generated through a fully adaptive greedy algorithm that recursively minimizes the $L^2$-error within a six-dimensional Tonnetz embedding of musical notes. The partitioning process employs a splitting criterion based on harmonic distance, resulting in regions that accurately reflect the intrinsic harmonic relationships among notes. 
The reconstructed music scores obtained through piecewise-constant functions and the mean values of the notes inside each wedge are used as a new simplified scores human-readable and playable. Some experiments on a corpus of symbolic music scores of three different composers
are performed to assess the proposed approach.
\end{abstract}

\medskip
\noindent\textbf{Keywords.} Symbolic music; Tonnetz; graph; wedge partitioning tree; computational musicology.

\medskip

\section{Introduction}
\label{sec:intro}

Computational methods for the music analysis can be categorized depending on the features to be extracted. Conventional approaches represent music as an \emph{audio} signal (e.g., a waveform, or a time-frequency representation such as a spectrogram), from which features are identified by signal-processing \cite{muller2015fundamentals} or deep-learning techniques \cite{goodfellow2016deep, purwins2019deepaudio}. Differently, the present paper is focused on the \textit{symbolic} domain, where the subject is the musical score transformed into a mathematical object, the \textit{graph}, explicitly encoding notes, rhythm, voices and texts, before any sound realization.
Computational analysis of symbolic music has experienced
a methodological renewal in the last decade~\cite{zhang2023symbolic,karystinaios2024graphmuse}, driven by deep learning \cite{briot2020deeplearning,purwins2019deepaudio} and music information retrieval \cite{muller2015fundamentals}. Symbolic formats such as MusicXML~\cite{good2001musicxml} and MIDI~\cite{midi1996} expose
explicit information on pitch, duration, voicing, lyrics, dynamics, and score position. Such granularity has motivated a sustained effort to design representations exploiting it without flattening it into
either piano-roll matrices~\cite{kong2020largescale,kim2020deep} or
linear token streams~\cite{huang2019musictransformer,fradet2021miditok} (see Appendix).

\noindent Among the proposed approaches, \emph{graph-based} representations have
emerged as particularly well-suited to the hierarchical, polyphonic, and temporal nature of a score. It has been used for cadence detection~\cite{karystinaios2022cadence}, Roman numeral
analysis (see Appendix)~\cite{karystinaios2023roman}, voice
separation~\cite{karystinaios2023voice}, expressive performance
rendering~\cite{jeong2019graph} and composer
classification~\cite{zhang2023symbolic}.
A particular instance has been proposed in the previous work of the
authors~\cite{licciardi2025rossini}, where a heterogeneous score-graph has been designed by integrating three types of node:\texttt{note\_voice} for the melodic
line, \texttt{syllable} for the lyrics, and \texttt{note\_piano} for the accompaniment; each one connected by seven relation types encoding: sequential (\texttt{voice-next-voice, syllable-next-syllable, piano-next-piano}),
vertical (\texttt{piano-vert-piano}) and cross-modal (\texttt{syllable-sung\_on\_head-voice, syllable-sung\_on-\quad \quad 
voice, piano-sung\_on-voice}) alignment. In the present paper the score-graph construction is adopted.

Music scores contain substantial structural redundancy, especially in the accompaniment, where harmonic figurations, Alberti basses, arpeggiated chords, and ostinati repeatedly express the same harmonic sequence over extended temporal spans (see Appendix). 
At the moment, in the symbolic domain, the compression of this redundancy has been approached through algorithmic
information theory~\cite{cilibrasi2005clustering}, sequence-level
redundancy reduction~\cite{shibata2025lzmidi}, wavelet transforms \cite{Francomano,chui1992wavelets,strang1996wavelets} on melodic signals~\cite{velarde2013waveletmelodic}, and
degradation toolkits motivated by transcription
robustness~\cite{mcleod2020mdtk}. None of these methods is simultaneously: i)~adaptive to the geometry of the underlying graph, ii)~aware of harmonic distance rather than mere chromatic
distance, and iii)~constrained to produce a reconstruction that
remains a valid musical score in the same scale and instrumentation as
the original.

Graph signal processing \cite{shuman2013emerging,ortega2018graphsignal} generalizes classical signal processing to data defined on graphs, where signals are associated with nodes and the underlying structure encodes their relationships between pairs. Recent advances in graph signal processing, as binary wedge partitioning trees and fully adaptive greedy (FA-greedy)minimization of the $L^2$-error~\cite{erb2023wedgelets,erb2025segmentation}, offer interesting tools to manipulate the complexity of information included in a graph and reduce the computational costs. In this paper, the same starting point is adopted and handled, together with the use of a multi-dimensional Tonnetz embedding of notes  \cite{harte2006detecting,cohn1998introduction}, in order to study a structure-preserving compression scheme for symbolic music scores and ensure the harmonic distance during the process maintaining the adaptiveness of the graph geometry.
The  paper is organized as follows: Section~\ref{sec:background} reviews the relevant state-of-the-art on graph-based representations of symbolic music and existing compression schemes; Section~\ref{sec:method} presents the adopted method on the structure-preserving compression scheme for score-graph, based on Tonnetz-driven graph wedgelet; in Section \ref{sec:experiments} music experimental setup is provided and Section~\ref{sec:conclusion} concludes and foreshadows the future works.

\section{Related Work on Graph-Based Music Representation and Compression Strategies}\label{sec:background}

The present section reviews the state-of-the-art relevant to the proposed approach, focusing on three main threads: i) graph-based models for music and ii) compression schemes for musical scores.

\begin{enumerate}[i)]
    \item \textbf{Graph-Based Representations of Symbolic Music}. 
The representation of music scores as graph structures has received increasing attention in recent years. For instance is the work of Jeong et
al.~\cite{jeong2019graph}, who modelled MusicXML scores as inputs to
a Gated Graph Neural Network, with notes as nodes and explicit edges
encoding sequential, simultaneity and same-voice relations, for expressive piano performance rendering. The approach has
been substantially refined and extended by Karystinaios, Widmer and
collaborators in a sequence of works targeting different analytical
tasks: cadence detection cast as node classification on a homogeneous
note graph~\cite{karystinaios2022cadence}; Roman numeral analysis with onset-wise predictions aggregated from note features~\cite{karystinaios2023roman}; voice separation cast as link
prediction on a heterogeneous graph with multiple temporal relation
types~\cite{karystinaios2023voice}, and more recently multi-task
music analysis with a shared graph
backbone~\cite{karystinaios2025analysisgnn}. The community-level
consolidation of this line of research is represented by the Python library "GraphMuse"~\cite{karystinaios2024graphmuse}, which standardizes
score-to-graph conversion and graph training on symbolic data.

In parallel, Zhang et al.~\cite{zhang2023symbolic} have provided the most thorough comparative evaluation of symbolic music representations to date, contrasting matrix-based (piano roll, see Appendix),
sequence-based, and graph-based representations across
three piece-level classification tasks: composer, performer and
difficulty. Their graph representation is homogeneous, with notes as
nodes and four edge types (onset, during, follow, silence), and is
processed with stacked GraphSAGE Neural Network \cite{hamilton2017graphsage}. Authors report that graph
representations are competitive with sequence-based ones, with
task-dependent advantages.

The extension to \emph{heterogeneous} graphs, where nodes of distinct types coexist, has been pursued primarily in two directions: a) in the
audio domain, song-level heterogeneous graphs with audio, lyric and
tag layers have been used for music emotion
recognition~\cite{liu2022hetmer} and artist
similarity~\cite{dasilva2024artist}; b) in the symbolic domain, the
heterogeneous score graph of Licciardi et
al.~\cite{licciardi2025rossini} introduces the constructions described in Section \ref{sec:intro}.

\item \textbf{Compression Schemes of Music Composition}. Compression of music composition has been addressed from several
distinct perspectives. Algorithmic information theory, particularly
the Lempel-Ziv family \cite{zivlempel1977}, has been used by Cilibrasi and
Vit\'anyi~\cite{cilibrasi2005clustering,cilibrasi2004algorithmic} to
cluster music files via the normalized compression distance. The
underlying compressor is general-purpose and treats the score as a
byte string, with no awareness of musical structure. A modern
revisitation of this idea, applied to symbolic music generation
rather than structure analysis, is the LZMidi
system~\cite{shibata2025lzmidi}. Wavelet-based methods have been
employed by Velarde et al.~\cite{velarde2013waveletmelodic} to
analyze monophonic pitch signals using the standard discrete Haar
transform \cite{chui1992wavelets}, treating melodies as one-dimensional time series of MIDI
integers; such construction is sequential and neglects the interactions induced by simultaneity and polyphony. The MIDI Degradation
Toolkit~\cite{mcleod2020mdtk} simulates transcription errors rather
than compressing in the strict sense, but it shares the spirit of
producing a controlled lossy version of a symbolic file. Finally, the rapidly growing literature on symbolic music
tokenisation~\cite{fradet2021miditok,huang2019musictransformer} can
be interpreted as a form of vocabulary-level compression, optimized for downstream neural processing rather than for reconstruction.

Within this landscape, existing approaches largely operate either at the signal level, sequential event level, or at the symbolic token level, without explicitly integrating polyphonic interactions with harmonic geometry in a single representation. The methodology introduced in Section~\ref{sec:method} involves both aspects within a unified modelling framework.
\end{enumerate}

\section{Graph Wedgelet Method for Score-Graph Reduction}\label{sec:method}

The work \cite{erb2023wedgelets} presents a wedgelet framework for graph signal compression. The main idea is to partition the graph into regions called \textit{wedges}, within which the signal is assumed to be constant. By representing the graph signal as a collection of these piecewise constant regions, the framework achieves an efficient compressed representation.
The wedges are  created using a binary graph partitioning  tree  built recursively splitting  the graph to minimize $L^{2}$-error. \\
In the paper, the partitioning strategy is applied to a score-graph ($S$) based on the Tonnetz embedding \cite{harte2006detecting,cohn1998introduction}, thus  preserving the harmonic structure. This tree is performed using the \emph{shortest-path distance}, which ensures that every wedge is a connected subgraph~\cite{erb2023wedgelets}. The  $L^{2}$-error in the partitioning is computed in a six-dimensional Tonnetz embedding of notes, whose euclidean metric captures \emph{harmonic} rather than chromatic proximity. Each wedge is represented by the mean of the corresponding region in the Tonnetz embedding, which is then quantized by assigning the nearest note in the Tonnetz, without introducing notes that are not present in the original score.

\subsection{Graph Wedgelets and Binary Wedge Partitioning Trees}
\label{sec:bg-bwp}

Let $G = (V, E)$ denote a finite, simple, undirected, connected graph
with $n = |V|$ vertices, equipped with a non-negative, symmetric
distance function $\dist : V \times V \to \mathbb{R}^{+}\cup \{0\}$. The
unweighted shortest-path distance, induced by the topology of $G$, is
the canonical choice and is the one adopted throughout the present
work. The construction of graph wedgelets, recently introduced in \cite{erb2023wedgelets}, consists of three fundamental steps: (i) wedge splitting, (ii) binary wedge partitioning (BWP) and (iii) wedgelet approximation. The first step concerns the definition of a wedge split.
\begin{definition}[Wedge split,
{\cite[Def.~IV.1]{erb2023wedgelets}}]\label{def:wedgesplit}
A dyadic partition $\{V', V''\}$ of a vertex set $V$ is a
\emph{wedge split} of $V$ if there exist two distinct nodes
$u, w \in V$ such that
\begin{equation}
V' = \{v \in V \mid \dist(v, u) \leq \dist(v, w)\}
\quad\text{and}\quad
V'' = \{v \in V \mid \dist(v, u) > \dist(v, w)\}.
\label{eq:wedgesplit}
\end{equation}

\end{definition}

When $\dist$ is the shortest-path distance on a connected graph,
$V'$ and $V''$ are themselves connected subgraphs of
$G$~\cite[Prop.~IV.2(3)]{erb2023wedgelets}. This property is
fundamental for the present application, in which every wedge is
required to be a musically contiguous fragment of the accompaniment.\\

Iterating wedge splits produces a binary tree of nested partitions of
$G$.

\begin{definition}[BWP{} tree,
{\cite[Def.~IV.3]{erb2023wedgelets}}]\label{def:bwp}
A \emph{binary wedge partitioning} (BWP{}) tree $\mathcal{T}_{V_M}$ of
$G$ with respect to the ordered set $V_M$ =  $\{ v_1, \ldots, v_M \}$$
\subseteq V $ is constructed recursively as follows.
\begin{enumerate}
\itemsep0pt
\item The root of  $\mathcal{T}_{V_M}$ is the set $V$, forming the
trivial partition $\mathcal{P}^{(1)} = \{V_{v_1}^{(1)}\} = \{V\}$
and associated to $v_1 \in V_M$.
\item For a partition $\mathcal{P}^{(m)} = \{V_{v_1}^{(m)}, \ldots,
V_{v_m}^{(m)}\}$ of  $ V$ in $\mathcal{T}_{V_M}$ associated to $v_i \in
V_{v_i}^{(m)}$, $i \in \{1, \ldots, m\}$, $m < M$, consider the
node $v_{m+1} \in V_{v_j}^{(m)}$ for some $j \in \{1, \ldots, m\}$.
The set $V_{v_j}^{(m)}$ is split by a wedge split based on $v_j$
and $v_{m+1}$ into the two disjoint sets $V_{(v_j,v_{m+1})}^{(m)+}$
(containing $v_j$) and $V_{(v_j,v_{m+1})}^{(m)-}$ (containing
$v_{m+1}$), yielding the new partition $\mathcal{P}^{(m+1)} =
\{V_{v_1}^{(m+1)}, \ldots, V_{v_{m+1}}^{(m+1)}\}$.
\end{enumerate}
\end{definition}
The final partition $\mathcal{P}^{(M)} = \{V_{v_1}^{(M)}, \ldots,
V_{v_M}^{(M)}\}$ is uniquely determined by the ordered set $\mathcal{T}_{V_M}$. \\

In the following the third step, regarding the wedglet approximation {\cite[Alg.~2]{erb2023wedgelets}, is detailed.\\

Given a graph signal $f : V \to \mathcal{X}$  in a normed vector  
space $(\mathcal{X}, \|\cdot\|)$, and  $\mathcal{P}^{(M)}$ the  partition generated by a BWP tree
$\mathcal{T}_{V_M}$, the \emph{wedgelet approximation of order} $M$ is~
\begin{equation}
\mathcal{W}_M f(v) \;=\; \sum_{i=1}^{M}
\bar{f}_{V_{v_i}^{(M)}}\, \omega_{v_i}^{(M)}(v),
\label{eq:wedgelet-approx}
\end{equation}
where 
\begin{equation}
\bar{f}_{V_{v_i}^{(M)}} \;=\;
\frac{1}{\bigl|V_{v_i}^{(M)}\bigr|}
\sum_{v \in V_{v_i}^{(M)}} f(v)
\label{eq:meanblock}
\end{equation}
and

\begin{equation}
\omega_{v_i}^{(M)}(v)=
\begin{cases}
1 & v \in V_{v_i}^{(M)}\\
0 & v \notin V_{v_i}^{(M)}.
\end{cases}
\end{equation}

The equality $\mathcal{W}_n f = f$ holds when $M=n$.\\

Three greedy procedures for selecting  $V_M$ in an $f$-adapted fashion are introduced in \cite{erb2023wedgelets}, where the FA-greedy variant is claimed to provide the best performance. In the following, the FA-greedy is summarized for the reader's convenience.  \\

At each refinement step $m$, two choices are
made.
\begin{enumerate}
\itemsep0pt
\item \emph{$V_{v_j}^{(m)}$ selection.} The region $V_{v_j}^{(m)}$ to be split
is chosen according to the rule~\cite[Eq.~(5)]{erb2023wedgelets}
\begin{equation}
j \;=\; \argmax_{i \in \{1, \ldots, m\}}\;
\bigl\| f - \bar{f}_{V_{v_i}^{(m)}} \bigr\|_2,
\label{eq:blockselection}
\end{equation}

\item \emph{ $v_{m+1}$ selection.} The vertex $v_{m+1}$ is chosen as
the node that minimises the squared $L^2$-error of the resulting
wedge split, i.e., $v_{m+1} \in V_{v_j}^{(m)}$ minimises
\begin{equation}
\bigl\| f - \bar{f}_{V_{(v_j,v)}^{(m)+}} \bigr\|^2_2
\;+\;
\bigl\| f - \bar{f}_{V_{(v_j,v)}^{(m)-}} \bigr\|^2_2
\label{eq:facriterion}
\end{equation}
over all candidates $v \in V_{v_j}^{(m)}$, in accordance
with~\cite[Eq.~(6)]{erb2023wedgelets}.
\end{enumerate}

In terms of applications, graph wedgelets have so far been applied
to grid graphs arising from
two-dimensional images, for compression~\cite{erb2023wedgelets} and
split-and-merge segmentation of biomedical
images~\cite{erb2025segmentation}, and to two small benchmark
non-grid graphs (the Minnesota road network and an Erd\H{o}s-R\'enyi
random graph \cite{erb2023wedgelets}).

\subsection{The Tonnetz and Angular Embedding}
\label{sec:bg-Tonnetz}

A pitch class is the equivalence class of all notes separated by one or more octaves.
From this point on, notes are identified with their corresponding pitch classes.
The harmonic distance between two pitch classes $p$ and $q$ is poorly captured
by the chromatic distance $|p - q| \bmod 12$ for $p, q \in
\mathbb{Z}/12\mathbb{Z}$. \\

A canonical counterexample is the pair
$(C, G)$: chromatic distance $5$ (or $7$, depending on direction),
yet harmonically the closest possible relation, the perfect fifth,
on which Western tonal syntax rests (see Appendix). A second example is the pair
$(C, D)$: chromatic distance $2$, yet harmonically remote, lying two
fifths away on the circle of fifths.

The \emph{Tonnetz}, originally introduced by
Euler~\cite{euler1739tentamen} as a two-dimensional lattice of
fifth- and third-related notes, developed into a planar tonal
network by Riemann and Oettingen \cite{oettingen1866,riemann1914ideen}, and recast into a torus geometry
by neo-Riemannian
theorists~\cite{cohn1998introduction,tymoczko2011geometry}, encodes this discrepancy: pitch classes related by a perfect fifth,
a major third, or a minor third are made adjacent. The
present work adopts a simple algebraic realisation of this idea, in
which each pitch class is mapped to one or more points on a circle in
such a way that harmonically related pitch classes are placed close
together.

Let
\begin{equation}
    v \in V = \{0, 1, \ldots, 127\} \equiv \mathbb{Z}_{128}
\end{equation} denote a
generic note in MIDI notation, and let
\begin{equation}
    p = v \bmod 12 \in \overline{V} = \{0,1,\ldots,11\} \equiv \mathbb{Z}_{12}
\end{equation}
be its pitch class. For a pitch class $p$, let $\theta_p = 2\pi p / 12$
denote its position on the chromatic circle. The na\"ive embedding
\[
p \mapsto (\cos\theta_p,\sin\theta_p).
\]
reproduces the chromatic
circle, and hence the same inadequate metric. The key device is to
rescale the angle by an integer factor $k$ before placing the pitch
class on the circle:
\begin{equation}
\varphi_k(p) \;=\; \bigl(\cos(k\theta_p),\, \sin(k\theta_p)\bigr)
\;\in\; \mathbb{R}^2 .
\label{eq:fourier-Tonnetz}
\end{equation}

The choice of $k$ selects which musical interval becomes the unit of proximity on the circle:

\begin{itemize}
\item $k=3$: coincidence occurs when $|p-q|\in\{0, 4, 8\}$, i.e.
for pitch classes related by a major third. This is the
\emph{major-third (augmented-triad) axis}.
\item $k = 4$: coincidence occurs when $|p-q| \in \{0, 3, 6, 9\}$, i.e.
for pitch classes related by a minor third. This is the
\emph{minor-third (diminished-seventh) axis}.
\item $k = 7$: since $7$ and $12$ have no common factor, no two
distinct pitch classes coincide, instead the twelve pitch classes are
spread around the circle in circle-of-fifths order, with fifth-related
pitch classes ($|p - q| \equiv 7 \pmod{12}$) placed at the minimal
angular distance $2\pi/12$. This is the \emph{circle-of-fifths axis}.
\end{itemize}
The six-dimensional Tonnetz embedding is obtained by concatenating
three such projections for a triple of pairwise distinct integers
$k_1, k_2, k_3$:
\begin{equation}
\Phi(p) =
\bigl(\cos(k_1\theta_p),\, \sin(k_1\theta_p),\,
       \cos(k_2\theta_p),\, \sin(k_2\theta_p),\,
       \cos(k_3\theta_p),\, \sin(k_3\theta_p)\bigr) \in \mathbb{R}^6 .
\label{eq:Tonnetz6d}
\end{equation}
Any permutation of
$k_1, k_2, k_3$ yields the same induced $L^2$ metric. For instance, if $V = \{21, 22, \ldots, 108\}$ then
$\Phi(p)$, with $p \in V$, are the embedded pitch classes of the piano range.

Related pitch-class representations within machine-learning pipelines
include the two-dimen\-sional Tonnetz-image representation of Chuan and
Herremans~\cite{chuan2018Tonnetz} for polyphonic-music model\-ling with
convolutional architectures, and the chord embedding of Kehoe et
al.~\cite{kehoe2020Tonnetz} for next-chord prediction via LSTMs \cite{choi2016textlstm,hochreiter1997lstm,licciardi2025partial,ala2023deep}. The
construction is also closely related to the \emph{tonal mean-based}
feature of Harte et al.~\cite{harte2006detecting} in audio chord
recognition. To the best of the authors' knowledge, no prior work
uses such a six-dimensional embedding as an error space for adaptive
partitioning of graph signals on a music graph.
Let  $f : V \to  \mathbb{R}^6$ the MIDI signal in Tonnetz embedding, the compression scheme proposed in the present work is composed by: \\
1. an \textbf{encoder}, which converts the piano subgraph $G_{piano}(S)$ together with its
MIDI signal $f$ into a representation of $M$ wedges (see Fig. \ref{fig:bwp});\\
2. a \textbf{ decoder}, which reconstructs from this representation a
piecewise-constant MIDI signal ${\mathcal{W}}_M f$  on the
same vertex set. 
\\
The effect of the full pipeline on a real excerpt from the corpus is
shown in Fig.~\ref{fig:toy}.

\begin{figure}[h!]
\centering
\includegraphics[width=\textwidth]{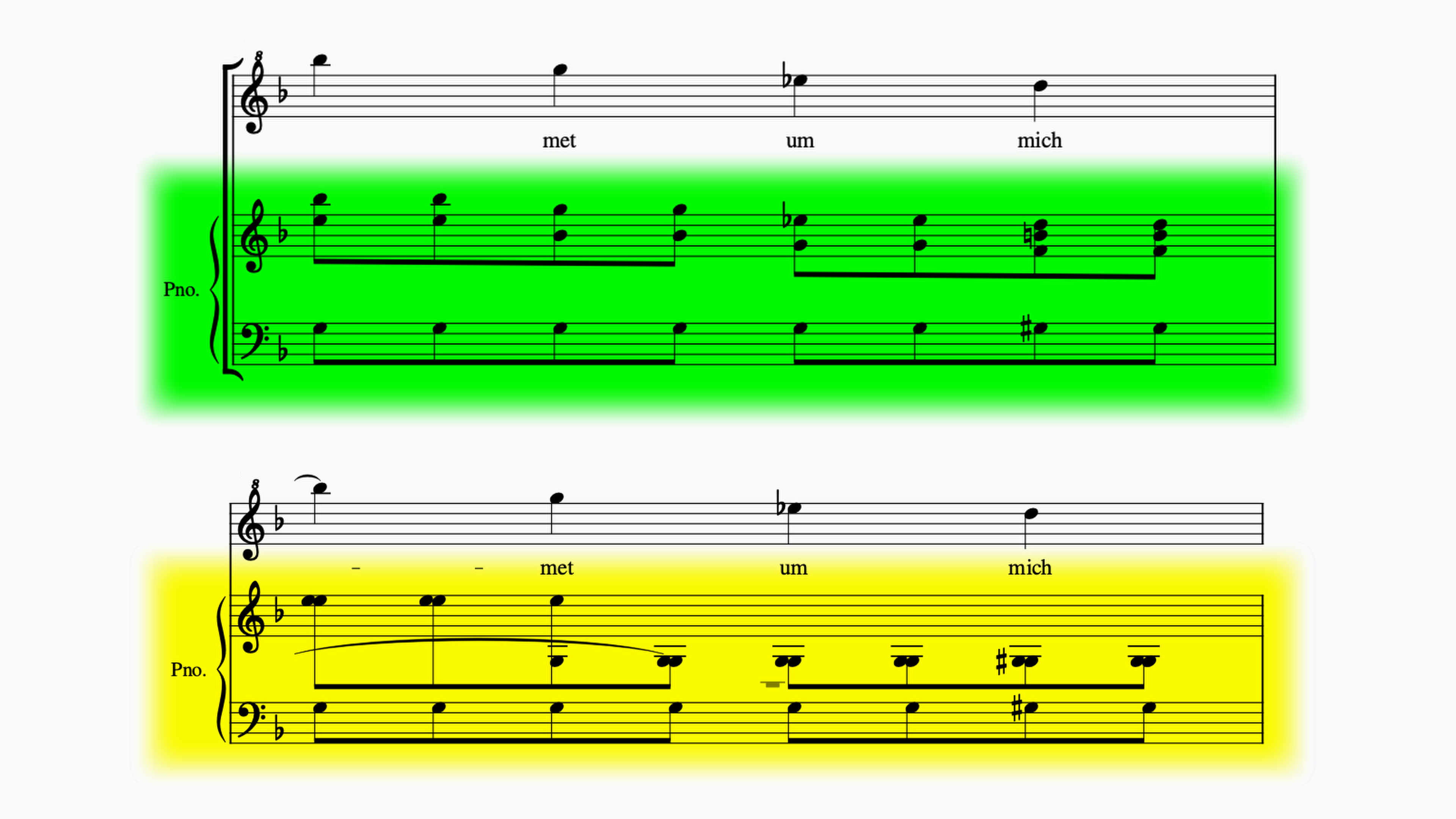}
\caption{W.~A.~Mozart, ``Der H\"olle Rache kocht in meinem Herzen''
(\emph{Die Zauberfl\"ote}, K.~620), measure~9: original (top, piano
in green) and compressed (bottom, piano in yellow).
The vocal line is untouched and the accompaniment is rendered by the mean-based-projected value of each wedge.}
\label{fig:toy}
\end{figure}

\subsection{Encoding and Decoding}
\label{sec:method-encoding}

The encoder applies the FA-greedy BWP{} construction to the Tonnetz-valued signal $f$,
with two specifications that fix free parameters of the original
algorithm to the music-symbolic setting:
\begin{itemize}
\itemsep0pt
\item \emph{Initial centre.} The initial node $v_1 \in
V_{\texttt{piano}}$ required to initialise the trivial partition
$\mathcal{P}^{(1)} = \{V_{\texttt{piano}}\}$ is taken as the first
piano vertex in temporal order. Alternative musically motivated
choices (e.g., the node whose Tonnetz embedding is closest to the
mean-based of $f$) can be done.
\item \emph{Wedge budget.} For a piano subgraph with
$N = |V_{\texttt{piano}}|$ vertices, the number of wedges is set to
\begin{equation}
M \;=\; \max\bigl(2,\, \lfloor r\, N \rfloor\bigr)
\label{eq:wedgebudget}
\end{equation}
where $r \in [0, 1]$ is a compression ratio that controls the
rate of the reduction.
\end{itemize}

\noindent The complete procedure for the encoder is summarized in Algorithm~\ref{alg:encoding} and an example on a real case is shown in Fig. \ref{fig:bwp}.

\begin{figure}[htbp]
 
\includegraphics[width=\textwidth]{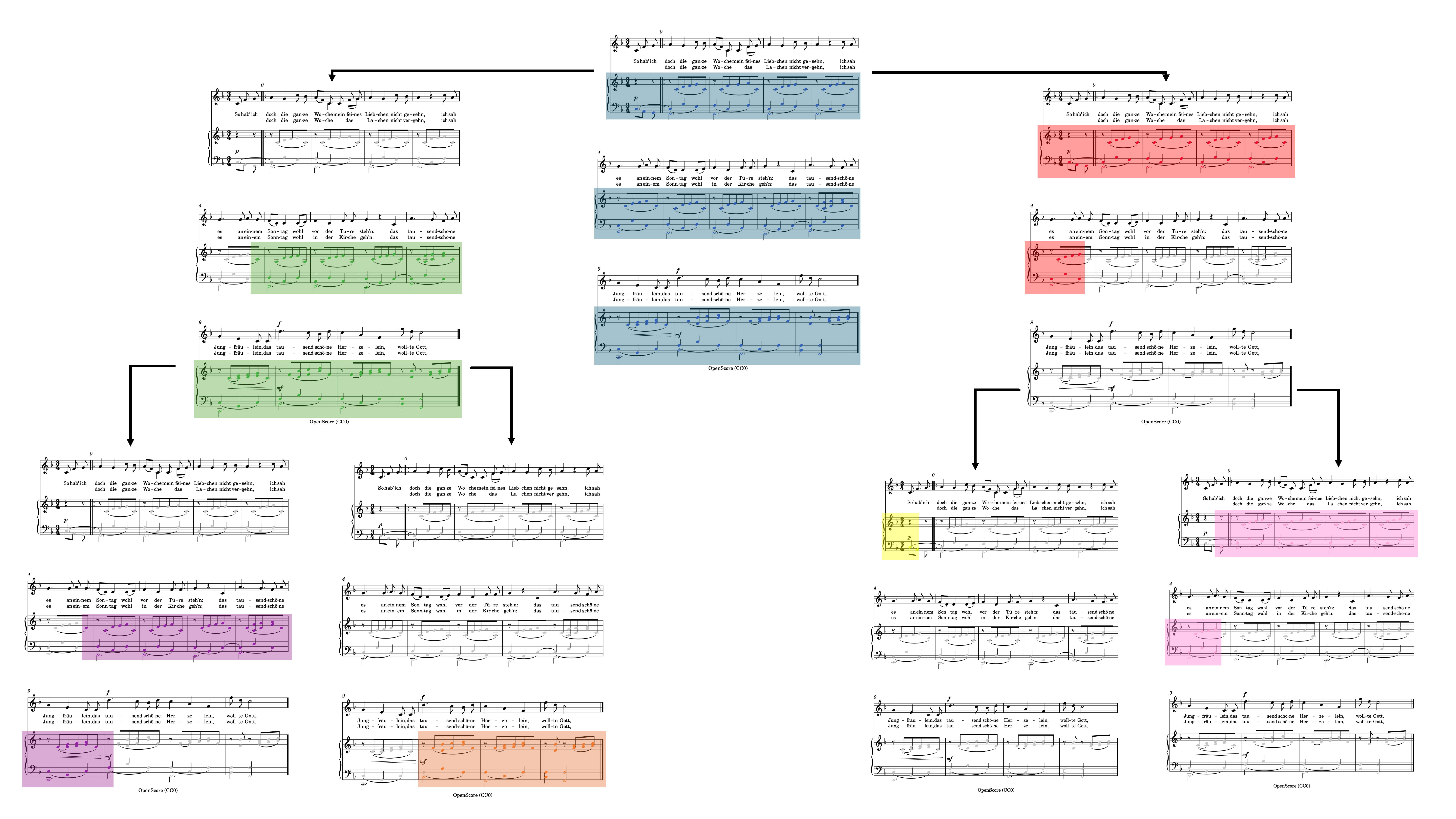}

  \caption{\small Binary wedge partitioning of the piano subgraph, shown on the score. For the selected example, $r = 0.03$ produces $M = 4$ final leaves. Starting from the root (the whole accompaniment), each node is split into two
  children by the FA-greedy criterion. Different notes assigned to different leaves of the tree are highlighted in distinct colors.}
  \label{fig:bwp}
\end{figure}

\begin{algorithm}[h!]
\caption{Tonnetz-driven FA-greedy BWP{} encoding of the piano signal.}
\label{alg:encoding}
\begin{algorithmic}[1]
\Require Piano subgraph $G_{piano}(S)$; MIDI signal $\midi$;
compression ratio $r \in [0,1]$.
\Ensure Ordered centres $V_M= \{
v_1, \ldots, v_M\}$.
\State Compute the pairwise shortest-path distances on $G_{piano}(S)$
\State Map each piano vertex to its Tonnetz vector
\Comment{(see eq. \eqref{eq:Tonnetz6d})}
\State Set the number of wedges $M \gets \max(2, \lfloor r N \rfloor)$
\State Take $v_1$ as the first piano vertex in temporal order, and
start with the whole vertex set as a single wedge
\For{$m = 2, \ldots, M$}
  \State Select the wedge whose vertices are worst approximated by a
  single Tonnetz value 
  \State Within it, pick the new center $v_m$ that makes the two
  resulting wedges most harmonically homogeneous 
  \State Split the wedge around its two centres, assigning each vertex
  to the nearer one 
\EndFor
\State For each final wedge store its centre of its vertices
\State \Return $V_M=$$\{\nu_1, \ldots, \nu_M\}$
\end{algorithmic}
\end{algorithm}

The decoder takes the set of ordered centers $V_M$ and locates the wedge containing $v_m$ and splits it around its centre. For each wedge $V_{v_i}$ computes the mean using Eq. \eqref{eq:meanblock} and the reconstruction described in Eq. \ref{eq:wedgelet-approx}. Finally, starting from ${\mathcal{W}}_Mf(v)$, in order to build a playable and readable music score, computes
\begin{equation}
    \widetilde{\mathcal{W}}_Mf{(p)} = \argmin_{\begin{array}{l}
   \scriptscriptstyle p_{j} \in \mathrm{PC}(V_{v_i}), \\[-0.1cm]
   \scriptscriptstyle j = 1, 2, ..., |V_{v_i}|
  \end{array}
   }
  \| \Phi(p_{j}) - {\mathcal{W}}_Mf{(v)} \|_2,
\end{equation}
where $\mathrm{PC}(V_{v_i})$ denotes the set of pitch classes present in the wedge.
The complete procedure for the decoder is summarized in Algorithm \ref{alg:decoding}.

This procedure ensures that every note belongs from a range of selected notes chosen by the composer so reducing the harmonic structure complexity without distorting it.

\begin{algorithm}[h!]
\caption{Mean-based decoding of the piano signal and musical embedding}
\label{alg:decoding}
\begin{algorithmic}[1]
\Require Piano subgraph $G_{piano}(S)$; shortest-path distances; Tonnetz signal
$f$; ordered centres $V_M = \{v_1, \ldots, v_M\}$.
\Ensure Reconstructed MIDI signal $\widetilde{\mathcal{W}}_Mf(p)$.
\State Start with the whole vertex set as a single wedge
\For{$m = 2, \ldots, M$}
  \State Locate the wedge containing $v_m$ and split it around its centre, by adopting the same distance
\EndFor
\Comment{the final partition $\mathcal{P}^{(M)} = \{V_{v_1}, \ldots, V_{v_M}\}$
is recovered}
\For{each wedge $V_{v_i} \in \mathcal{P}^{(M)}$}
  \State Compute the mean-based $\bar f_{V_{v_i}}$
  \State Compute $\mathcal{W}_M f(v)$
  \State Select $\widetilde{\mathcal{W}}_Mf{(p)} = \argmin_{\begin{array}{l}
   \scriptscriptstyle p_{j} \in \mathrm{PC}(V_{v_i}), \\[-0.1cm]
   \scriptscriptstyle j = 1, 2, ..., |V_{v_i}|
  \end{array}
   }
  \| \Phi(p_{j}) - {\mathcal{W}}_Mf{(v)} \|_2$ 
\EndFor
\Comment{musical embedding}
\State \Return $\widetilde{\mathcal{W}}_Mf(p)$
\end{algorithmic}
\end{algorithm}

\subsection{Export Pipeline to MusicXML}\label{sec:method-export}

To produce a playable and human-readable
compressed score, the reference implementation interfaces with the
\textsf{music21} library~\cite{cuthbert2010music21} as follows. The
original MusicXML file is parsed into a \textsf{music21} score
object, and the sequence of \texttt{Note} and \texttt{Chord} elements
of the piano part is extracted in onset-time order. For each
event, the MIDI pitch is overwritten with the corresponding
component of $\widetilde{\mathcal{W}}_Mf$, clipped to the valid
MIDI range. All other notational attributes (duration,
metric position, voicing, articulation, dynamics, lyrics, beaming,
and stem direction) are left unchanged. The modified score is then
written to a new MusicXML file.

\section{Simulation Setup}\label{sec:experiments}

In this section some experiments  on a corpus of 70
symbolic scores of three different composers in MusicXML format are performed. For each score, the piano subgraph $G_{piano}(S)$ and its MIDI
signal are extracted as described in
Section~\ref{sec:bg-Tonnetz}, yielding, per each score, a graph of
$N = |V_{\texttt{piano}}|$ piano vertices on which the compression
operates.
Let $f$ be the Tonnetz-valued piano signal
(Section~\ref{sec:bg-Tonnetz}), and $\mathcal{W}_M f$ its
wedgelet approximation of order $M = \max(2, \lfloor r N \rfloor)$,
which assigns to each vertex the mean-based $\bar f_{V_{v_i}}$ of the
wedge $V_{v_i}$ containing it (Eq.~\eqref{eq:wedgelet-approx}). The \emph{root-mean-square error} between the original
signal and its wedgelet approximation, is computed vertex-wise in the
embedding space:
\begin{equation}
\mathrm{RMSE} \;=\;
\sqrt{\,\frac{1}{N} \sum_{v \in V_{\texttt{piano}}}
\bigl\| f(v) - \mathcal{W}_M f(v) \bigr\|_2^2 \,}.
\label{eq:rmse}
\end{equation}
The squared distances are averaged over the $N$
piano vertices (so that scores of different length are comparable) and the metric measures the average harmonic standard deviation introduced, per note, by approximating the piano signal with $M$ wedge-constant values.

For each score in the corpus, the wedgelet approximation is computed for
each attainable compression ratio $r$, and the corresponding $\mathrm{RMSE}$ is
recorded. The per-score errors are then aggregated across the corpus as
mean and standard deviation (std) at each ratio $r$, yielding the curve of Fig.~\ref{fig:errorcurve}. When $r = 1.0$
($M = N$) exact reconstruction is obtained, whereas as $r \to 0$ ($M \to 2$) the
approximation reaches the coarsest admissible level.

\begin{figure}[htbp]
\centering
\includegraphics[width=0.5\textwidth]{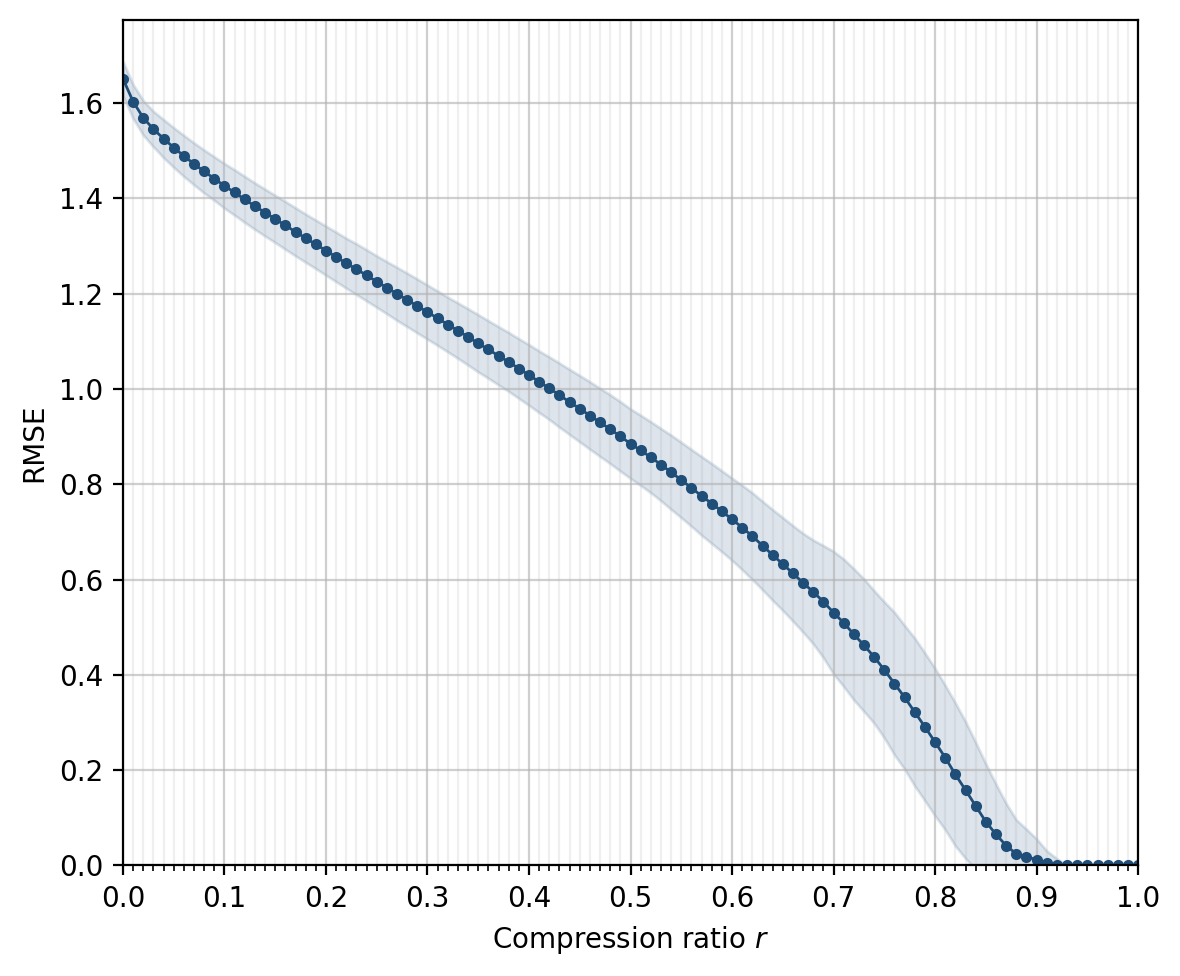}
\caption{RMSE and std of the wedgelet approximation in the Tonnetz embedding versus the compression ratio $r \in [0.0, 1.0]$.}
\label{fig:errorcurve}
\end{figure}

To make the effect of compression directly visible on the score itself,
Fig.~\ref{fig:compression-levels} shows a single measure of an
accompaniment reconstructed at the different compression ratios
$r\in\{0.2*i: i=0,1,...,5\}$. 
In each $r$, a
notehead is drawn in \emph{green} when the compressed reconstruction
preserves the original pitch of that note, and in \emph{yellow} when the
note has been reassigned to its wedge. The proportion of
yellow noteheads thus measures, visually and per note, how much the score
has been altered at that level: at $r = 1.0$ the reconstruction is
identical to the original (all green), at the coarsest level almost
every note is reassigned (predominantly yellow), and the intermediate
$r$ display the gradual transition between the two extremes.

\begin{figure}[h!]
\centering
\includegraphics[width=\textwidth]{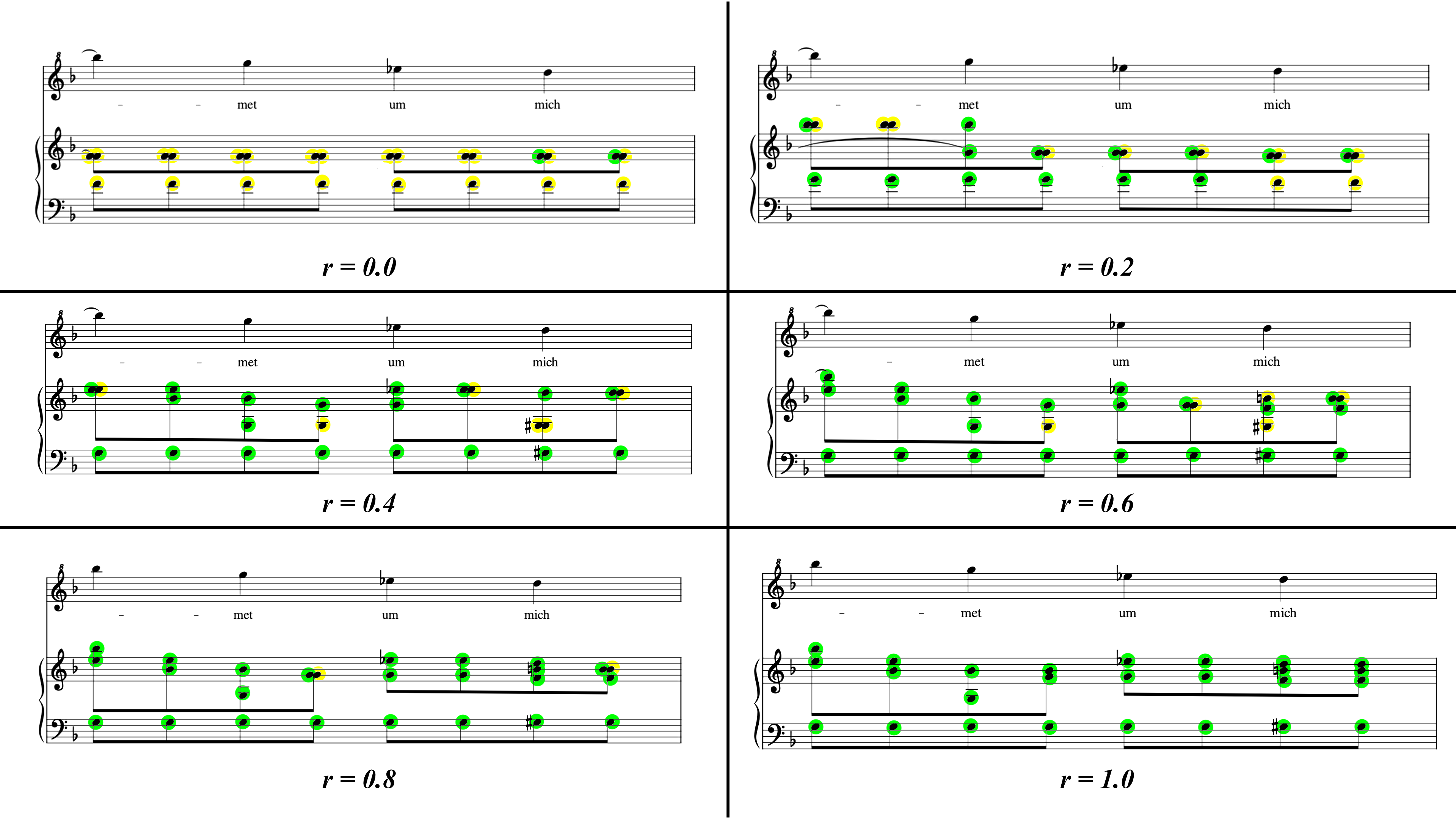}
\caption{One measure of an accompaniment reconstructed at compression
ratios $r\in\{0.2*i: i=0,1,...,5\}$. Green noteheads:
pitches preserved by the compression; yellow noteheads: pitches reassigned
to the wedge. The fraction of yellow noteheads decreases as
$r$ grows, visualising the fidelity-compression trade-off.}
\label{fig:compression-levels}
\end{figure}

\section{Conclusion and Future Work}\label{sec:conclusion}
The present work introduces an adaptive compression scheme for
symbolic music scores, extending the consolidated graph wedgelet framework.
Two musically motivated modifications are proposed. The first consists of a six-dimensional Tonnetz embedding integrated into the FA-greedy BWP algorithm, allowing the partitioning criterion to reflect harmonic rather than chromatic proximity. The second employs a mean-based decoder followed by quantisation onto the pitch classes present in each wedge so ensuring  that the reconstructed pitches remain on the original pitch grid while preventing the introduction of pitch classes absent from the input and that every note belongs from a range of selected notes chosen by the composer so reducing the harmonic structure complexity without distorting it. Several research directions remain open.
The present scheme compresses the homogeneous piano
subgraph in isolation; extending the wedge split and the error metric
across node types would allow the adaptive partitioning of the full
\emph{heterogeneous} score graph, in which pitched and non-pitched
elements (melody, lyrics) coexist and require type-aware notions of
distance and harmonic embedding.
At the moment the reconstruction assigns a single constant value to each
wedge; replacing the per-wedge constant with other functions could capture intra-wedge harmonic motion
that a single value cannot represent.
Finally, the broader aim is to develop a software tool for the automatic generation of simplified musical arrangements. The present work establishes a necessary condition for achieving this objective by ensuring that a compressed arrangement remains recognisable as the original piece from which it was derived.

\section{Appendix}
\label{app:glossary}

For the convenience of readers this appendix
collects extended definitions of the music-theoretic \cite{piston1987harmony} and
representational notions used in this work. 

\begin{description}

\item[Pitch class] One of the twelve categories into which all pitches
are grouped under \emph{octave equivalence}: two pitches whose
frequencies differ belong to
the same pitch class. Formally the twelve pitch classes form the cyclic
group $\mathbb{Z}/12\mathbb{Z} = \{0, 1, \ldots, 11\}$, with $0 = C$,
$1 = C\sharp$, and so on. A pitch class discards octave information and
retains only the chromatic identity of a note; the embedding $\Phi$ of
this work is defined on pitch classes because harmonic
function is octave-invariant as shown in Fig. \ref{fig:pitchclass}.

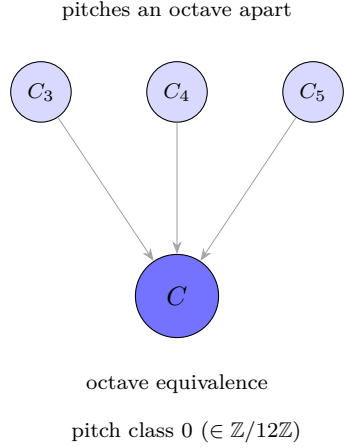
\begin{figure}[h!]
\centering
\begin{tikzpicture}[scale=0.9]
  \foreach \oct/\lab in {0/{$C_3$},1/{$C_4$},2/{$C_5$}}{
    \node[draw,circle,fill=blue!15,minimum size=8mm,font=\scriptsize] (c\oct) at (\oct*2,2) {\lab};
  }
  \node[draw,circle,fill=blue!55,minimum size=11mm,font=\small] (pc) at (2,-1) {$C$};
  \foreach \oct in {0,1,2}{ \draw[->,>=Stealth,gray!70] (c\oct) -- (pc); }
  \node[font=\scriptsize,anchor=west] at (0.3,-3){pitch class $0$ ($\in\mathbb{Z}/12\mathbb{Z}$)};
  \node[font=\scriptsize,anchor=south] at (2,2.9){pitches an octave apart};
  \node[font=\scriptsize,anchor=north] at (2,-2){octave equivalence};
\end{tikzpicture}
\caption{Octave equivalence: pitches separated by one or more octaves
($C_3, C_4, C_5, \ldots$) collapse to a single pitch class, an element
of $\mathbb{Z}/12\mathbb{Z}$. The embedding $\Phi$ is defined on pitch
classes, discarding octave information.}
\label{fig:pitchclass}
\end{figure}

\item[Piano-roll] A two-dimensional binary (or velocity-valued) matrix
representation of a score in which the columns index discrete time
steps and the rows index the $128$ MIDI pitches. A nonzero entry at
position $(t, p)$ records that pitch $p$ is sounding at time step $t$.
The representation is regular and grid-structured as shown in Fig. \ref{fig:pianoroll}.

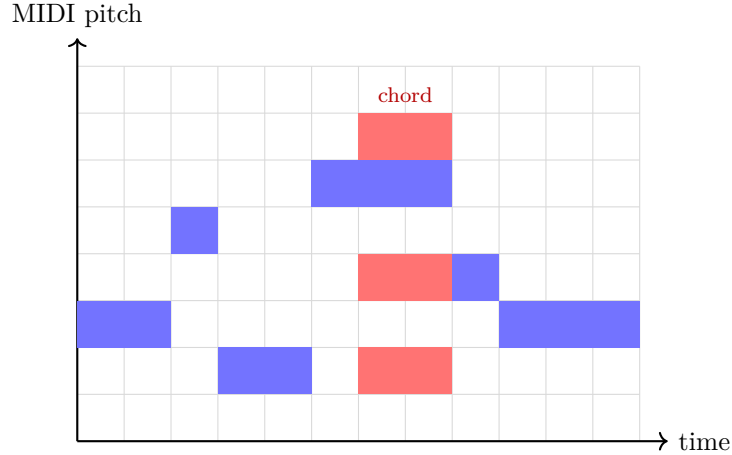
\begin{figure}[h!]
\centering
\begin{tikzpicture}[scale=0.62]
  \def\T{12}\def\P{8}
  \foreach \x in {0,...,\T} \draw[gray!30] (\x,0)--(\x,\P);
  \foreach \y in {0,...,\P} \draw[gray!30] (0,\y)--(\T,\y);
  \draw[->,thick] (0,0)--(\T+0.6,0) node[right]{\small time};
  \draw[->,thick] (0,0)--(0,\P+0.6) node[above]{\small MIDI pitch};
  \fill[blue!55] (0,2) rectangle (2,3);
  \fill[blue!55] (2,4) rectangle (3,5);
  \fill[blue!55] (3,1) rectangle (5,2);
  \fill[blue!55] (5,5) rectangle (8,6);
  \fill[blue!55] (8,3) rectangle (9,4);
  \fill[blue!55] (9,2) rectangle (12,3);
  \fill[red!55] (6,1) rectangle (8,2);
  \fill[red!55] (6,3) rectangle (8,4);
  \fill[red!55] (6,6) rectangle (8,7);
  \node[font=\scriptsize,red!70!black] at (7,7.4){chord};
\end{tikzpicture}
\caption{Piano-roll representation: columns index time steps, rows index
MIDI pitches. Each rectangle is a sounding note, its width its duration;
three notes sharing a column (red) form a chord.}
\label{fig:pianoroll}
\end{figure}

\item[Token stream] An encoding of a score as a one-dimensional,
ordered sequence of discrete symbols (\emph{tokens}), each denoting a
musical event or attribute, for instance a note onset, a pitch, a
duration, a bar line, or a time-shift. Token streams are the symbolic
analogue of text and are the standard input to sequence models such as
recurrent networks and transformers. The encoding linearises the score,
which captures sequential dependencies efficiently but represents
simultaneity (chords, polyphony) only implicitly, through the ordering
and grouping of tokens. For instance, a single bar containing a quarter-note $C_4$ followed by a
half-note $\{C_4, E_4, G_4\}$ chord may be linearised, in a MIDI-like
tokenisation, as the stream
\begin{center}
\ttfamily\small
Bar\_1 \textbar{} Position\_0 \textbar{} Pitch\_60 \textbar{} Duration\_4 \textbar{}
Position\_4 \textbar{} Pitch\_60 \textbar{} Pitch\_64 \textbar{} Pitch\_67 \textbar{} Duration\_8
\end{center}
where \texttt{Pitch\_60} denotes middle $C$ ($C_4$), simultaneous pitches
sharing a \texttt{Position} encode a chord, and \texttt{Duration} values
are expressed in sixteenth-note units. The score is thus flattened into a
one-dimensional symbol sequence.

\item[Roman numeral analysis] A standard system of harmonic analysis
that labels each chord of a composition by its \emph{function} relative
to the local key, rather than by its absolute pitch content. Each chord
is denoted by a Roman numeral indicating the scale degree of its root
(e.g.\ \textnormal{I} for the triad built on the first degree of the
scale, \textnormal{IV} on the fourth, \textnormal{V} on the fifth),
with additional symbols encoding chord quality and inversion. Because
the labelling is relative to the key, the same numeral identifies the
same harmonic function across different keys, making the analysis
transposition-invariant; this is the property that motivates its use as
a learning target in key-independent harmonic models.

\item[Cross-modal relations] In the heterogeneous
graph of this work, a relation is called \emph{cross-modal} when it connects
vertices of \emph{different} node types (for instance an edge linking
a note vertex to a lyric vertex) as opposed to a
\emph{within-type} relation connecting vertices of the same type (e.g.\
two note vertices). Cross-modal relations let information flow between
heterogeneous elements of the score (pitched material, text, structural
markers) and are what distinguishes a heterogeneous score graph from a
homogeneous note graph.

\item[Alberti bass] A stock keyboard accompaniment figure in which the three
notes of a triad are not struck simultaneously but broken into a
repeating arpeggiated pattern, conventionally in the order
lowest-highest-middle-highest as shown in Fig. \ref{fig:alberti}. The device sustains the harmony while
generating continuous rhythmic motion in the left hand, and is
pervasive in Classical-era piano writing. For the present work it is a case of harmonic
redundancy: a long stretch of notes that, although melodically varied,
spells out a single underlying chord and is therefore highly
compressible in a harmony-aware metric.

\begin{figure}[h!]
\centering
\includegraphics[width=0.7\textwidth]{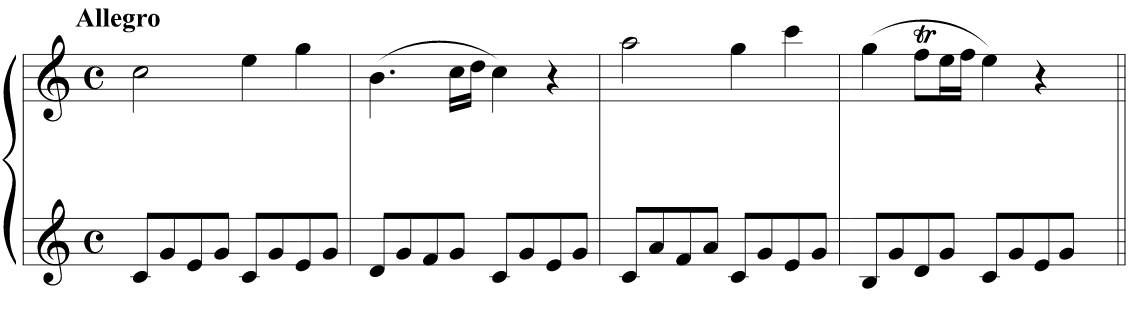}
\caption{An Alberti bass figure: the notes of a single triad are not
struck together but broken into the repeating pattern
lowest-highest-middle-highest. Although the surface is melodically
active, every note belongs to the same underlying harmony, making the
passage strongly redundant and, in a harmony-aware metric, highly
compressible.}
\label{fig:alberti}
\end{figure}

\item[Ostinato] A figure (melodic, rhythmic, or harmonic) that is
repeated persistently throughout a passage while other voices change
around it. In an accompaniment role the ostinato repeats a fixed
pattern with little or no alteration of its pitch and temporal content,
producing strong local redundancy as shown in Fig. \ref{fig:ostinato}. As with the Alberti bass, an
ostinato accompaniment is a setting in which a wedgelet approximation
is expected to be efficient, since many consecutive notes are
harmonically equivalent.

\begin{figure}[h!]
\centering
\includegraphics[width=0.7\textwidth]{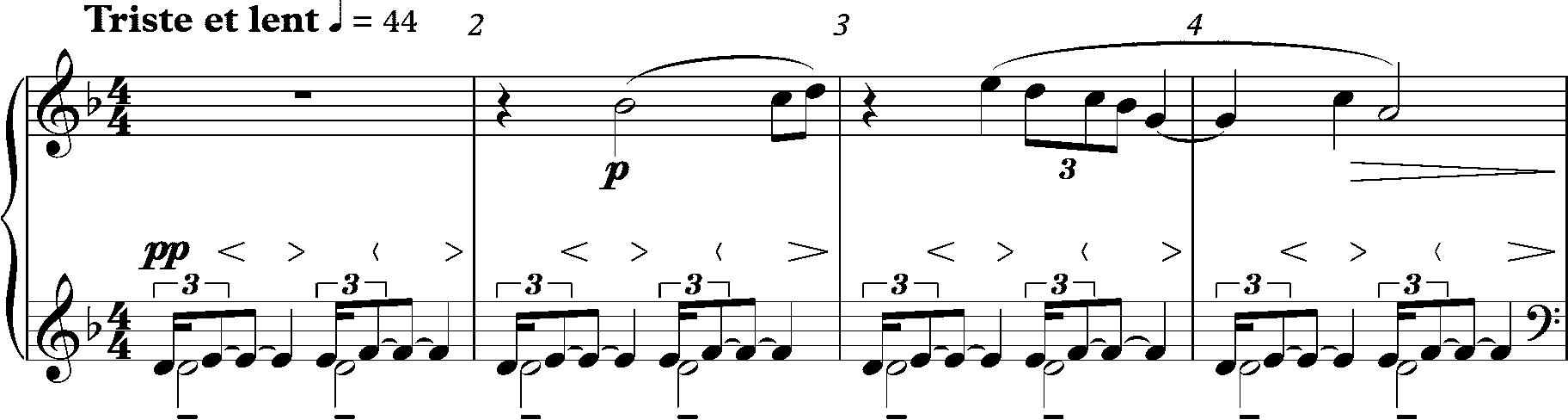}
\caption{An ostinato accompaniment: a fixed pattern repeated with little
or no change in its pitch and rhythmic content. The persistent
repetition produces strong local redundancy across consecutive notes,
which a wedgelet approximation can capture with few wedges.}
\label{fig:ostinato}
\end{figure}

\item[Chromatic distance] The separation, measured in semitones,
between two pitch classes $p, q \in \mathbb{Z}/12\mathbb{Z}$, given by
$|p - q| \bmod 12$ on the cyclic group of the twelve pitch classes. It
is the most immediate notion of pitch proximity, but it does \emph{not}
reflect \emph{harmonic} proximity, i.e.\ the perceived closeness of two
notes within tonal syntax. The canonical illustration is the pair
$(C, G)$: a perfect fifth apart and harmonically about as close as two
distinct pitch classes can be, yet at chromatic distance $7$; conversely
$(C, D)$, harmonically remote, lies at chromatic distance $2$. This
mismatch is the central motivation for replacing the chromatic metric
with the Tonnetz embedding of Section~\ref{sec:bg-Tonnetz}.

\item[Perfect fifth] The interval between two pitch classes separated
by seven semitones (chromatic distance $7$), or equivalently one step
along the circle of fifths. It is, after the octave, the most consonant
interval and the primary structural interval of Western tonal
harmony, the relation on which keys, triads, and the circle of fifths
are built. Its harmonic centrality, despite its large chromatic
distance, is exactly what a harmony-aware metric must capture and a
chromatic one cannot as shown in Fig. \ref{fig:chromatic}.

\begin{figure}[h!]
\centering
\begin{tikzpicture}[scale=1.0]
  \def\R{2.6}
  \foreach \i/\name in {0/C,1/{C\#},2/D,3/{D\#},4/E,5/F,6/{F\#},7/G,8/{G\#},9/A,10/{A\#},11/B}{
    \node[circle,draw,fill=gray!10,minimum size=7mm,inner sep=0pt,font=\scriptsize]
      (n\i) at ({90-\i*30}:\R) {\name};
  }
  \draw[->,>=Stealth,green!55!black,line width=1.2pt] (n0) to[bend right=18] (n7);
  \draw[->,>=Stealth,red!75!black,line width=1.2pt]   (n0) to[bend left=30]  (n2);
  \node[font=\scriptsize,align=left,anchor=west] at (-3.0,-3.5)
    {\textcolor{green!45!black}{$C\!\to\!G$: perfect fifth, harmonically close, \emph{chromatic} $7$}\\
     \textcolor{red!75!black}{$C\!\to\!D$: harmonically remote, \emph{chromatic} $2$}};
\end{tikzpicture}
\caption{The twelve pitch classes on the chromatic circle. Chromatic
distance (semitone separation) misrepresents harmonic proximity: the
perfect fifth $C$--$G$ (green) is harmonically close yet at chromatic
distance $7$, whereas $C$--$D$ (red) is harmonically remote yet at
chromatic distance $2$. This mismatch motivates the Tonnetz embedding.}
\label{fig:chromatic}
\end{figure}
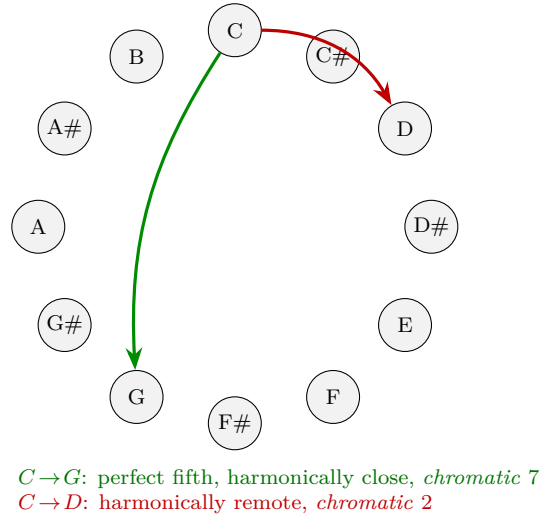

\end{description}

\section*{Acknowledgement}
E.C., E.F. and S.L. are supported by GNCS-INdAM 2026 project “Metodi polinomiali e kernel per l’approssimazione da dati discreti e integrali con software OS” (CUP E53C25002010001) and by UMI-TAA research group.

\end{document}